\documentclass{elsart}
\usepackage{epsfig}
\usepackage{subfigure}
\usepackage{amssymb}
\journal{Nucl. Instr. and Meth. in Phys. Res. A}
\begin{document}

%
\begin{frontmatter}

 \title{Development of a compact scintillator hodoscope with
	wavelength-shifting fibre read-out}
 \author{C. P. Achenbach\corauthref{cor}}
 \corauth[cor]{Corresponding author. Present address: Institut f{\"u}r 
	Kernphysik, Joh.\ Gutenberg-Universit{\"a}t Mainz, 
	J J Becher-Weg 45, 55099 Mainz, Germany.
	Tel.: +49--6131--3925831; fax: +49--6131--3922964.}
 \ead{patrick@kph.uni-mainz.de}
        and \author{J. H. Cobb} 
 \address{University of Oxford, Sub-department of Particle Physics, 
	Denys Wilkinson Bld., Keble Rd., Oxford, OX1 3RH, UK}

 \begin{abstract} 

  We report on the prototyping of a plastic scintillator hodoscope
  with wavelength-shifting fibre read-out by a multi-anode
  photomultiplier as part of the development of a detector for cosmic
  ray muons to be carried aboard an aircraft.  Light yield and light
  attenuation measurements on single- and double-clad
  wavelength-shifting fibres were performed. Low power, low-threshold,
  discriminators were designed. A prototype 16-channel hodoscope with
  two planes was built and tested with cosmic rays. After correcting
  for geometrical factors a global intrinsic efficiency of $\epsilon >
  98$\,\% was obtained in both planes. The overall performance of the
  hodoscope proved it to be well suited for the {\sc Adler} experiment
  to measure the high altitude muon flux.

 \end{abstract}
  

\end{frontmatter}
%

\section{Introduction}
We report on the development of a plastic scintillator hodoscope with
wavelength-shifting (WLS) fibre read-out by a multi-anode
photomultiplier (PMT). The work described in this paper was part of
the research and development activity for the {\sc Adler} experiment
(\underline{A}irborne \underline{D}etector for \underline{L}ow
\underline{E}nergy \underline{R}ays)~\cite{Adler2001}. The object of
the {\sc Adler} experiment is to measure the flux of low energy
atmospheric muons at aircraft cruising altitudes as a function of
geomagnetic coordinates.  Observations of muons at ground level are of
limited value because of the large and unknown amount of kinetic
energy (typically 2\,GeV) lost by the muons in the atmosphere. While
ground measurements are widely reported in the literature, there have
been very few attempts to measure the muon flux as a function of
altitude and latitude. The effect of the geomagnetic field is
recognised as very important in cosmic ray cascade
calculations. Accordingly, the proposed measurement could provide an
important calibration of the calculations of atmospheric neutrino
fluxes performed by the Bartol
group~\cite{Barr1989},\cite{Agrawal1996}, and by Honda, Kajita,
Kasahara and Midorikawa~\cite{Honda1990}. Accurate predictions of the
neutrino fluxes are needed in the analysis of data from underground
experiments such as Super-Kamiokande~\cite{Fukuda1998},
MACRO~\cite{Ambrosio1998} and Soudan~2~\cite{Allison1998} to obtain a
better understanding of flavour oscillations of muon neutrinos.

Section~2 describes the design criteria for the hodoscope which follow
from the special measurement conditions of the proposed {\sc Adler}
experiment. Simple estimates which guided the design of the hodoscopes
and front-end electronics are given in Section~3. The light yield and
light attenuation measurements on single- and double-clad WLS fibres
which were performed to optimise the design of the hodoscopes and
electronics are described in Section~4. A prototype 16-channel
hodoscope with associated discriminators boards was constructed and
tested to verify the performance of the design. In Section~5 we
present the results of cosmic ray tests of this hodoscope.

\section{Design Considerations}
The aim of the initial phase of the {\sc Adler} experiment was to
design a compact detector which could be carried by an aircraft to
measure the muon flux at altitudes close to the production maximum at
different geomagnetic locations. Severe constraints on mass, size,
power consumption and efficiency of the instrument are imposed by the
requirement that it could be carried by a commercial aircraft. The
whole apparatus must be certified as meeting rigorous airworthiness
requirements (specified in ``Eurocae ED-14D''). For example, it must
be designed to withstand forward accelerations of up to 9g and lateral
accelerations of 3g.

The conceptual {\sc Adler} detector consists of scintillator
hodoscopes interleaved with a passive lead absorber followed by an
active scintillator absorber surrounded by a veto counter.  The
individual counters must be highly efficient for an absolute
measurement of the atmospheric muon count rate. Furthermore, the
detector as a whole must provide sufficient spatial resolution to
correlate hits in a number of hodoscope planes and to assign them to
unique muon tracks. The inner components of the detector are shown in
Fig.~\ref{Fig:adler_sketch}. The active area of each hodoscope was
chosen to be $A \approx 800$\,cm$^2$.  Muons which stop in the
absorber are distinguished by the observation of delayed coincidences
between signals from absorber and the hodoscopes and no activity in
the surrounding veto-counters.

The {\sc Adler} detector requires of the order of 100 individual
hodoscope channels. The use of conventional scintillation counters and
single channel PMTs would be costly in terms of space and weight. It
was therefore decided to base the design on the use of scintillators
with WLS readout in order to achieve a compact detector. The WLS
fibres serve the dual purpose of primary light collection and
subsequent flexible light transmission over a distance of 50 --
100\,cm. The drawback of the use of WLS fibres for primary light
collection is that the overall `light yield' (photo-electrons per unit
energy deposit) is low compared with conventionally instrumented
scintillation counters and it was necessary to undertake an
optimisation study to ensure that efficient hodoscopes could be built.

The electronic read-out of the hodoscopes also had to be economical in
terms of cost, space and power consumption. A main consideration was
the avoidance of high power consuming electronic crates and
modules. In addition, the use of multi-anode photomultiplier tubes
greatly reduces the number of high-voltage (HV) power supplies and
voltage dividers required. A compact tube like the Hamamatsu
Photonics~\cite{Hamamatsu} R5900-00-M16 multi-anode photomultiplier
(M16) fitted very well into the over-all scheme.  The performance of
the M16 tube was first reported in~\cite{Yoshizawa1997} and
comprehensive tests on gain, uniformity, cross-talk and behaviour in
magnetic fields have been conducted by~\cite{Enkelmann1998}. Since
this study, Hamamatsu has continued to make improvements to the tube
particularly in quantum efficiency and collection efficiency of the
dynode chain. The M16 has a 0.8\,mm thick borosilicate window spaced
only 2\,mm from the first dynode, a single $17.5 \times 17.5$\,mm$^2$
bialkali or S10 photocathode, individual metal channel dynodes and the
output signals are obtained from independent anodes. The tube can be
operated at negative HVs for dc-coupled anode read-out. It has a
typical gain of about $1-3\times 10^6$ when operated at $-$800\,V. The
$4 \times 4$\,mm$^2$ pixels are arranged in a $4 \times 4$ matrix and
separated by gaps of 0.275\,mm. The gain variation between pixels can
be up to a factor of three. Despite this potential difficulty, its
compact geometry of only 22 $\times$ 30 $\times$ 30\,mm$^3$ made it
very attractive for our application. Six M16 PMTs are sufficient to
perform the experiment.

\section{Design Estimates}
\subsection{Light Yield}
A minimum ionising particle typically loses $\sim$\,1.8\,MeV of energy
per cm. of plastic scintillator. The energy deposited is converted to
(blue) scintillation photons with a typical efficiency of $\epsilon_s \sim
1.5 \times 10^4$ photons/MeV. These photons will be absorbed
efficiently and re-radiated at a longer wavelength (green) if they
strike a WLS fibre. Since the size of a fibre is small compared with
the dimensions of a scintillator bar, the probability that a blue
photon will hit the fibre in one traversal of a bar is small, $P_W
\sim d/s$, where $d$ is the diameter of the fibre and $s$ is the
projected width of the scintillator bar. Blue photons may also be
absorbed with a probability $P_R = 1 - R$ when they strike the wall of
the scintillator bar, where $R$ is the reflection coefficient, or by
bulk absorption with a probability $P_A$ in the
scintillator. Therefore the probability that the ultimate fate of a
blue photon is to be absorbed by the WLS fibre is $\epsilon_W =
P_W/(P_W + P_R + P_A)$. Choosing $d=1.2$\,mm, $s=55$\,mm
(corresponding roughly to a bar with a 20\,mm $\times$ 35\,mm
cross-section) and assuming $1-R = 0.04 $ and $P_A = 0.02$,
$\epsilon_W$ evaluates to $\sim 0.24$.

The green photons are re-radiated isotropically in the WLS fibre and
only a small fraction, $\epsilon_t$ (typically 0.03 in a single clad
fibre), are captured within the acceptance cone of the fibre and
transmitted by internal reflection. The quantum efficiency,
$\epsilon_Q$, of a bialkali photocathode is $\sim$ 0.1 for green
photons. Assuming no further loss of light, {\it e.g.\/} due to
absorption in the fibre or losses at optical couplings, the light
yield of a system using 20\,mm $\times$ 20\,mm scintillator bars and
1.2\,mm diameter fibres is expected to be $Y_l = \epsilon_s \epsilon_W
\epsilon_t \epsilon_Q \approx 1.5 \times 10^4 \times 0.24 \times 0.03
\times 0.1 = 11$ photo-electrons/MeV. The average minimum ionising
particle traversing a 2\,cm bar perpendicularly would therefore
produce a signal of $\approx$40 photo-electrons, corner-clipping
particles giving proportionately less. This value of $Y_l$ is
sufficient to ensure that any inefficiency due to statistical
fluctuations of the number of photo-electrons is negligible.  More
careful estimates were made using a Monte Carlo (MC) programme which
tracked blue photons from the point of production by a muon to
eventual photo-electron production and correctly included all
geometrical and loss factors. The results of the MC programme
indicated essentially the same light yield. In particular, the result
of $\epsilon_W \sim 0.37$ confirmed the above estimate. Other
processes such as imperfect optical coupling and light loss due to the
bending of the WLS fibres would reduce the light yield. The latter
process was the subject of a separate study~\cite{Achenbach2003} and
indicated that a radius of curvature to fibre radius ratio of greater
than 65 results in a light loss of less than 10\,\% with the loss
occurring in a transition region at bending angles $\Phi \sim
\pi/8$\,rad.

\subsection{Discriminators}
The requirement of minimum power consumption excluded the use of fast
digitisation of the PMT signals; instead the signals would be
digitised by simple discriminators. It would be advantageous to use a
common threshold for all channels. Once allowance is made for
the gain difference between the PMT pixels and the requirement of
efficiency for minimum ionising particles, the discriminator
thresholds must be equivalent to a few photo-electrons, but greater
than one photo-electron to eliminate PMT dark noise and single photons
from cross-talk. Since the signals from the PMTs were expected to be
small and the power consumption of fast pre-amplifiers was
unacceptable, the design required the analogue signals from the PMTs
to be detected by low-threshold discriminators without amplification.
With a PMT gain of 10$^6$ a single photo-electron would develop a
3.5\,mV signal if integrated directly on 47\,pF.  Commercial
comparators can operate with threshold voltages as low as 5 -- 10\,mV
if proper precautions are taken to eliminate extraneous noise.


A further consideration for the design of the electronics was that the
full photo-electron signal equivalent to the estimated light yield is
not available unless the signal is fully integrated before
discrimination.  The time distribution of the photo-electrons is
determined by the decay time of the flours in the WLS fibre (7 --
10\,nS) and the collection time of the blue photons, estimated to be
another 5 -- 10\,nS in this geometry. An integration time of $\approx$
100\,nS or greater is required to achieve the full signal. This
requirement conflicts with the detector requirement that the
discriminators are able to re-trigger after 200\,nS for a wide range
of signals. With these considerations in mind, low-threshold
discriminators, mounted directly on the base of the photomultipliers,
were developed.


\section{Light Yield and Light Attenuation Measurements}
\subsection{Light Yield Measurements}
The study of light yield was performed using single- and double-clad
1.2\,mm diameter Bicron BCF-91A WLS fibres embedded in bars of Bicron
BC-400~\cite{Bicron} polyvinyltoluene-based plastic scintillator
(refractive index $n = 1.58$). The BCF fibres have a polystyrene core
of refractive index $n_{\it core}=$ 1.6. The single-clad fibres
(BCF-91A SC) have a thin polymethylmethacrylate (PMMA) cladding of
refractive index $n_{\it clad}=$ 1.49, the double-clad fibres (BCF-91A
DC) have an inner PMMA cladding and an outer fluorinated
polymethacrylate cladding of refractive index $n^\prime_{\it clad}=$
1.42. Double-clad fibres offer a significant improvement over
single-clad fibres since a larger fraction of light is trapped in the
fibre core. The 280\,mm long scintillation bars had a cross section of
$20 \times 35$\,mm$^2$. The fibres were located inside $1.5 - 2.0$\,mm
deep and 1.5\,mm wide grooves which had been filled with the optical
clear epoxy Bicron BC-600. The groove depth exceeded the fibre
diameter by 0.5 -- 0.8\,mm. Once the glue was set, the fibre ends were
polished and the scintillator bars painted with Bicron BC-620, a
diffuse white reflector composed mainly of titanium dioxide, TiO$_2$.


The WLS fibres were taken to an M16 and positioned by guidance holes
in a block of Tufnol high pressure laminate fitting the outer
dimensions of the PMT. Each fibre was centred on the centre of a
single photocathode pixel. The fibres were glued into the holes using
a fast epoxy, and coupled optically to the photocathode using optical
grease. The anode signals from the M16 were amplified using an
amplifier with a 10\,nS integration time constant and a gain of
$\sim$6. The output signals from the amplifier were discriminated and
counted. The M16 was operated with an HV of $-$850\,Volts. The
counting rate was measured as a function of threshold voltage when the
bars were illuminated with a weak (2k\,Bq) $^{106}$Ru $\beta$ source.
The maximum energy of the electrons emitted by $^{106}$Ru is 3.54\,MeV
(78\,\%) with sub-dominant contributions of transitions with 2.4\,MeV
(11\,\%) and 3.0\,MeV (8\,\%) maximum energy and additional gamma
lines.

The anode counting rate, which varied between 1000\,Hz and 100\,Hz,
was measured for two different channels as a function of discriminator
threshold from 30 to 300\,mV. The total count rate at minimum
threshold was normalised to accommodate gain variations between
pixels.  The peak of the single photo-electron signal was determined
by measuring the counting rate--threshold curve with no source but
with a controlled light leak, thereby enabling the threshold voltage
to be converted to equivalent photo-electrons.

The measured counting rate--threshold curves were compared with the
predictions of a MC simulation based on {\sc Geant}
3.21~\cite{GEANT}. In the simulation the energy depositions and
particle tracks were recorded in the active and inactive
materials. The assumption was made that the photo-electron output was
proportional to the energy deposited in the scintillator, {\it i.e.\/}
$N_{\rm pe}= Y_l\, E_{\rm dep}$ where $Y_l$ was the conversion factor
(`light yield') to be determined. The experimental photo-electron
resolution of the photomultiplier was simulated by a convolution of
the Poisson distribution, $P(n_{pe}|N_{pe})$, for the number of
photo-electrons released at the photocathode with a Gaussian
distribution with a width of $\sigma_E= \sigma \sqrt{n_{p.e.}}$ where
$\sigma\approx 40\,\%$ was deduced from single photo-electron
measurements with the M16. A match of the slopes of the simulated and
measured counting rate curves was achieved with a light yield, $Y_l$,
of 8\,photo-electrons$/$MeV of deposited energy as shown in
Fig.~\ref{Fig:light-yield}. Allowing for the attenuation in the WLS
fibre ($T\sim$ 55\,\%), the measured light yield was consistent with
the estimate of 11 photo-electrons/MeV given in Section~3.

Measurements also showed that a white diffuse reflector (BC-620 paint)
at the open end of a fibre allowed the collection of secondary photons
propagating away from the PMT and led to an increase in light yield of
$\sim$\,20\,\%.  A further increase was sought by using an aluminised
Mylar (a polyester film) foil, which would give specular reflection,
but no significant change was observed. A possible explanation could
be a misalignment of the reflector foil with respect to the fibre
axis. These results are shown in Table~\ref{tab:reflectors}.

We concluded that the light yield of a such a scintillator -- WLS
fibre combination is adequate for an efficient detector and that it
will allow the use of low power electronics.

\subsection{Light Attenuation Measurements}
Light attenuation in the WLS fibres was measured by illuminating
individual, loose fibres with a light-emitting diode. The LED, which
had a peak wavelength of $420$\,nm, was placed so as to shine
perpendicularly onto a WLS fibre at a variable distance from the end
coupled to a PMT. The LED was fixed to a housing which could move
smoothly over the tested fibres. Initially, the far end of the WLS
fibres were covered with a black absorber to remove all secondary
light. The green photons were detected by a 2~inch Philips XP-2230
photomultiplier tube. The average PMT anode current was determined by
measuring the voltage drop across 10\,M$\Omega$ in parallel with
12\,$\mu$F.  The intensity of illumination of the fibre was controlled
by varying the current through the LED. The PMT output varied linearly
with LED current over most of the working range, with non-linearities
at the low and high ends due to the non-linearity of the LED and
saturation (droop) of the PMT output current respectively.

As well as measuring the PMT anode current the counting rate of
discriminated PMT output pulses was also measured. An extremely good
correlation between count rate and anode current was measured because
the incoherent illumination of the fibre by the LED produced a
succession of single photo-electrons at the PMT cathode. Both anode
current and counting rate were used as a measure of the total light
intensity.

It was necessary to turn on the LED one hour before any measurements
to allow the system to become fully stabilised. After this time the
variations in anode current for fixed illumination did not exceed
2\,\%. All the detector components were located in a box providing
light tightness and mechanical protection. After exposure of the
photomultipliers to ambient light, such as fluorescent room lighting
when changing the experimental set-up, a temporary increase in anode
dark current was observed. The recovery time of the photomultiplier
tubes was of the order of 20 -- 60 minutes, depending on the
exposure. After such a waiting period a series of readings were taken
for each measurement. As a cross-check of any long term effects, the
light box was sealed on several occasions for 16 hours before turning
on the HV of the PMTs.

The light intensities transmitted over distances of up to 3\,m were
measured for single- and double-clad
fibres. Fig.~\ref{Fig:attenuation} shows the dependence of the light
intensity (dc anode current) as a function of photocathode distance
from excitation source, $d$, for different fibres.  The data shown in
Fig.~\ref{Fig:attenuation} demonstrates that the simple assumption of
a single attenuation length is incorrect.  The measurements are best
described by a sum of two exponential functions: $I_{ph}= I_L\,
e^{-d/\lambda_L} + I_S\, e^{-d/\lambda_S}$, where $\lambda_L$
represents a longer attenuation length, $\lambda_S$ a shorter one and
$I_{L/S}$ are the relative intensities.  We observed that the short
attenuation length dominated the light losses for distances $d <$ 1\,m
while the longer attenuation length was appropriate for distances of
$d \approx 1-5$\,m or more. Results for best fits on the measured
intensities are shown in Table~\ref{tab:attenuation}.

Most light detected within the first 50\,cm from the light source is
referred to as cladding light, which is not trapped in the fibre core
but by total internal reflections at the cladding-air interface. It is
usually a factor 3--4 more intense than core light. The transmission
of cladding light is strongly affected by the surface quality of the
fibre: cracks in the cladding or defects in the surface can cause
significant light losses leading to differences in the light yield of
otherwise identical fibres. Due to the unavoidable degradation of the
fibre surface, cladding light is expected to be attenuated more
strongly than core light. Since cladding light is attenuated by an
unpredictable amount, variations of its contributions to the total
light intensity are expected. To achieve an understanding of the
transmission of light in the core of WLS fibres, the removal of the
cladding light is useful. This was accomplished by coating the fibre
with an extra-mural absorber ({\it EMA}). The second set of data in
Table~\ref{tab:attenuation} shows the results of these
measurements. Since the LED housing prohibited the covering of the
first 5\,cm of fibre, the fits to the data points have been
constrained by $I_S= R\, I_L$, where $R$ is the ratio of light
intensities at $d \simeq 5$\,cm as given by the first set of
measurements. This method allowed the estimation of the relative
intensities of the two light components at zero fibre length. For
double clad fibres a higher $I_L/I_S$ ratio was observed which could
be explained by a lower proportion of light propagating by reflections
at the exposed surface. Errors on the attenuation lengths include an
estimated systematic error taking into account uncertainties due to
possible drifts of the output signals and the effects of
irregularities in the fibres. The results on the long attenuation
length were less accurate because the {\it EMA} was only applied to a
length of $d \approx$ 1\,m of fibre. Within the accuracy of the
measurements it can be concluded that the short attenuation length
decreased by around 30\,\%, whereas the long attenuation length was
not significantly affected by the addition of the {\it EMA}.

The improvement in trapping efficiency and attenuation length of
double-clad fibre with respect to single-clad fibre can be seen in
Fig.~\ref{Fig:attenuation}. For fixed illumination by the LED at
$d$=3\,m (where the cladding light has been substantially attenuated)
the PMT output currents are in the ratio of $\sim$3.8:2.0 ($=$1.9) for
the double- and single-clad fibres respectively.


{
\section{Prototype Hodoscope}
\subsection{Hodoscope Construction}
A prototype two-plane hodoscope was built to verify the design.  The
planes and scintillator bars had the same dimensions as those foreseen
for the full detector, but a somewhat different mechanical arrangement
for ease of access and testing. The two planes were orthogonal and
each contained eight scintillator bars of {20$\times$35 mm$^2$}
cross-section and 280\,mm long. The whole was enclosed in a single
rigid aluminium box approximately 600\,mm square.

The scintillator strips were mounted on either side of a 3\,mm thick
central aluminium sheet in a box consisting of a `picture frame' made
of $50 \times 8$\,mm aluminium bars. The central sheet slotted into
grooves machined in the bars. This method of construction provided
considerable rigidity for relatively little weight. The top and bottom
of the box were closed with thin aluminium covers. An M16 PMT and
discriminator cards, mounted directly on the PMT base, were located in
a demountable external housing.

A 1.2\,mm diameter WLS fibre was embedded to a depth of $\sim 1.8$\,mm
in each scintillator bar. Double-clad fibres (Bicron BCF-91A) were
used for the top hodoscope and single-clad fibres for the bottom
hodoscope. On leaving the scintillator bars the fibres were loosely
enclosed in opaque sleeving to prevent optical cross-talk and brought
to a Tufnol `cookie' on the side of the hodoscope box. Holes in the
cookie aligned with the pixels of the PMT.

A photograph of the aluminium frame with the top hodoscope plane and
the fibre routing is shown in Fig.~\ref{Fig:photograph}. The bottom
hodoscope plane is hidden below the central sheet.

\subsection{Electronics}
Signal discrimination was provided by low threshold, high speed
circuits. Two boards carrying eight channels of discriminators were
mounted directly onto the PMT base which also carried the divider
chain for the PMT dynodes. The circuit diagram of one channel is shown
in Fig.~\ref{Fig:discr}. The discriminators used Maxim 903
comparators. These have separate power supply connections for the
analogue input and digital output stages which is an advantage if the
lowest possible thresholds are to be achieved.  Simple $RC$
integration of the PMT anode signal is provided at the discriminator
input, the time constant of 10\,nS being chosen to match the estimated
light-collection time of scintillator -- WLS fibre combination. The
relatively large value of 47\,pF was chosen for the input capacitance
in order to reduce capacitative cross-talk. Independent externally
adjustable thresholds were provided in common for even and odd
channels. The $220$\,k$\Omega$ feedback resistor provided $5$\,mV of
hysteresis. The lowest practical threshold which could be achieved was
$\sim$6\,mV at the discriminator input.

Since the PMT signal due to many photo-electrons from the
scintillators arrives over a time of $\sim$ 10\,nS, the $R_{in}C_{in}$
integration time constant of 10\,nS at the input of the discriminators
provides only partial integration. The input signal to the
discriminators will depend non-linearly on the number of
photo-electrons. An isolated photo-electron will produce an input
signal of $V_{in}(1_{\rm pe}) = e\, G/C_{in}$ (where $e$ is the
electron charge and $G$ is the PMT gain) whereas the input voltage
from many ($N_{pe}$) photo-electrons will tend to $V_{in}(N_{\rm pe})
= N_{\rm pe}\, e\, G \exp{(-1)}/C_{in}$.

The TTL output stages of the discriminators were connected to drivers
which transmitted the Low Voltage Differential Signals (LVDS) to an
external interface board via 34-way `flat-'n-twist' cables. The same
cables were also used to provide the analogue and digital power, test
signals and the common thresholds to the discriminators.

A state-machine programmed `MACH 435' PGA chip managed the trigger
logic on the interface board and also performed handshaking with a PC
via a general purpose PCI card. The logic was designed to latch the
signals from the M16 on receipt of a trigger. Triggers could be
derived from external scintillation counters or internally by OR-ing
signals from the M16. Since the trigger signal from the external
conventional scintillation counters was late with respect to signals
from the M16 (due to the transit times in the PMTs), the discriminator
signals were stretched by up to 200\,nS, in 25\,nS steps in the PGA.

A LabView Virtual Instrument (VI) application running on the PC was
used for on-line analysis of the data and to write the data to files
for further analysis. The data acquisition scheme for the hodoscope is
shown in Fig.~\ref{Fig:dio}. The data samples read by the PC consisted
of bit patterns of 16 channels, each channel representing a
corresponding scintillator. These bit patterns were used for the
two-dimensional track reconstruction and for the determination of the
acceptance and efficiency of the hodoscope.

The hodoscope was mounted into a cosmic ray test stand. Two (three)
scintillator paddles of size 584 $\times$ 216\,mm$^2$, directly
coupled to Philips XP-2230 photomultipliers, were located above and
below the hodoscope. In addition, two lead shields, a larger block of
5\,cm height and 41\,cm length and a smaller one of 2.5\,cm height and
63\,cm length, were stacked between the scintillator paddles. A
schematic drawing is shown in Fig.~\ref{Fig:set-up}. Minimum ionising
muons lost a minimum of $\Delta E \approx 100$\,MeV kinetic energy in
the absorbers. A cosmic ray trigger was derived from the overlap
coincidence of the signals.

\subsection{Geometric Factor and Intrinsic Efficiency}
The cosmic ray trigger had an area a few times greater than the active
area of the hodoscope and accepted particles over a wide range of
angles. The geometrical efficiency of the top (bottom) hodoscope plane
was approximately 25 (12)\,\% for the two scintillator trigger and 20
(14)\,\% for the three scintillator trigger. It was obvious from the
orientation of the hodoscope strips (see Fig.~\ref{Fig:set-up}) that
the top plane scintillators accepted particles over a wider range of
angles. A MC simulation of the low energy cosmic ray flux was used to
determine the intrinsic efficiency of the scintillator strips and
light yield for minimum ionising muons.

The acceptance, $\alpha$, of the hodoscope for detecting an incident
muon was factored into two parts, the intrinsic efficiency,
$\epsilon$, and the geometrical factor, $(dA\, d\Omega)$. The
acceptance is simply the product of the two. The intrinsic efficiency
is the ratio of particles detected by the hodoscope to the number of
particles incident on area $dA$ from solid angle element $d\Omega$.
The geometrical factor was deduced from the simulation. The
differential and integral spectra of cosmic rays were taken
from~\cite{Grieder2001} and~\cite{Allkofer1971b}. A minimum momentum
cut of 0.2\,GeV/$c$ was used to allow for the two lead shields. In the
simulation, muons were generated on a virtual sphere enclosing the
hodoscope-trigger counter assembly. The points on the virtual sphere
were distributed so that any particle within an element of solid
angle, $d\Omega$, traversed in a downward sense a horizontal element
of area $dA$.  The zenith angle dependence of the flux was taken to be
$I(\theta) = I(0^\circ) \cos^2(\theta)$.  Azimuthal symmetry was
assumed.

The HV applied to the photomultiplier and the threshold on the
discriminator cards were chosen to achieve a high intrinsic efficiency
and low cross-talk. The plateau curves for thresholds of 45-55\,mV
showed a minimum slope at around $-$850\,V, where the counting rate
was the least sensitive to drifts in the HV.

The geometric factor could be eliminated by requiring a coincident hit
in one hodoscope planes to analyse the other. These global
efficiencies for each hodoscope plane, defined as $\epsilon_{T(B)} =
\Sigma T (B) / ( T_{\rm ext} \cdot\Sigma B (T) )$, are shown in
Fig.~\ref{Fig:hv_scan} as a function of HV and threshold voltage.  The
plateau voltage for the top plane, which used double-clad WLS fibres,
was $\sim$50\,V lower than for the bottom plane, consistent with the
greater light-yield expected for double-clad fibres. The efficiency of
each plane was determined at a high voltage of $-$850\,V,
corresponding to a PMT gain of 3$\times$10$^6$, and varying threshold
voltages. The measured acceptances and global efficiencies are given
in Table~\ref{tab:efficiencies}.  The average efficiency was $\epsilon
\approx$ 98.6\,\% for a threshold of 45 mV, corresponding to 10--12
photo-electrons.


Due to the geometry of the scintillator paddles the illumination of
the hodoscope by cosmic ray muons was non-uniform.
Fig.~\ref{Fig:acceptance} shows the relative acceptance, that is the
fraction of hits per trigger in each scintillator bar, for two sets of
measurements and the corresponding simulation values. The small
asymmetry in direction of the paddles' axes is due to their asymmetric
set-up. In the perpendicular direction the limited accuracy in the
positioning of the paddles led to a small left-right asymmetry.  To
check the efficiency of the two outermost scintillator bars, which
exhibited low count rates because of their small geometric acceptance,
the hodoscope was moved 5\,cm in either direction. The relative
acceptance of the two counters is shown in Fig.~\ref{Fig:acceptance}
by squares.

The global efficiencies, as defined above, include muons which clip
the corners of the hodoscope strips. To deduce the true intrinsic
efficiency for muons passing through at least 2\,cm of scintillator in
each plane, the MC simulation was used to generate a photo-electron
spectrum for the cosmic ray flux. The simulation included the gaps
between the scintillators due to their individual paint coating and
their finite thickness. Fig.~\ref{Fig:simulation}(a) shows the
simulated distribution of the number of photo-electrons spectrum for
the scintillator bars. There is a significant tail at low numbers of
photo-electrons due to the corner clipping muons. An estimate of the
minimum light yield of the hodoscope was possible from the measured
efficiencies $\epsilon \approx$ 98\,\%.  Fig.~\ref{Fig:simulation}(b)
shows a calculation of the global efficiencies, including
corner-clipping muons, versus threshold, for different light
yields. Since, after allowance for the difference in gain between
pixels, the largest effective threshold was 10--12 photo-electrons, we
conclude from Fig.~\ref{Fig:simulation}(b) that the light yield of the
hodoscope elements was larger than 8\,photo-electrons$/$MeV, in good
agreement with the estimates given in Section~3. This translates into
an absolute light yield of $N >$ 28\,photo-electrons for a normally
incident minimum ionising muon with an average energy deposition of
3.5\,MeV.

\subsection{Multiplicity}
At an operating HV of $-$850\,V and a threshold of 45\,mV, the hit
multiplicities were measured to be $\langle N \rangle_{\it TOP}= 1.13$
and $\langle N \rangle_{\it BOT}= 1.10$ for the three scintillator
trigger, compared with $\langle N \rangle_{\it TOP}= 1.01$ and
$\langle N \rangle_{\it BOT}= 1.03$ expected from the simulation. The
$\sim$10\,\% excess multiplicity was attributed to cross-talk in the
PMT and some electronics cross-talk. Cross-talk in the PMT can be
optical, which occurs due to the divergence of light as it passes
through the borosilicate glass window, or electronic, which occurs if
electrons begin the multiplication process in the wrong dynode
chain. We attributed the proportion of hits in non-neighbouring
channels ($\sim 1$\,\%) to random coincidences with background
radiation.

\section{Summary and Discussion}
A scintillator hodoscope has been developed to demonstrate a possible
detector concept for the measurement of the low energy muon flux in an
aircraft. The proposed detector ({\sc Adler}) had to be
self-contained, as small and light as possible, consistent with the
design objective and consume minimum electrical power. Our solution is
based on bars of plastic scintillation counters read out via WLS
fibres and multi-anode photomultiplier tubes. The multi-anode PMT
Hamamatsu R5900-00-M16 requires little power, is compact in size, and
has stable properties that make it suitable for our use with
scintillating fibres. In order to optimise the fibre routings inside
the hodoscope, the light attenuation characteristics of WLS fibres,
parameterised by two different attenuation lengths, have been
measured. Extra-mural absorbers applied to the fibre surface were
successfully used to measure the amount of cladding and core light in
the fibres. Simple, low power, low-threshold, discriminators were
developed. Two boards with eight discriminators each were directly
mounted on the the base of the photomultiplier tube, which also
carried the divider chain for the PMT dynodes.

A prototype scintillator hodoscope was constructed and tested with
cosmic rays. Data acquisition was performed without standard
electronic crates and modules, but with a single PGA chip and a high
speed digital I/O board of a PC. The intrinsic efficiency of the
hodoscope was found to be above 98\,\%. Average hit multiplicities per
plane were found to be $\sim$10\,\% higher than expected, suggesting
the presence of some optical and electronic cross-talk. The inferred
light yield of $>$ 8 photo-electrons/MeV was entirely adequate for the
intended purpose of the detector.

\section*{Acknowledgements}
We thank the MINOS group at the University of Oxford for lending us
the M16 and two voltage divider boards. We are also much indebted to
the technical staff, in particular to M.~Dawson, for their
assistance. We would like to thank K.~Ruddick of the University of
Minnesota for the programme to estimate light yields in scintillators.

This research was supported by the UK Particle Physics and
Astronomy Research Council (PPARC).


\clearpage
\newpage


%
\begin{table}[tbp]
  \begin{center}
    \caption{ Relative counting rates of scintillator -- WLS fibre
	combinations measured with a simple discriminator and scaler
	set-up. Diffuse (paint) and specular (aluminised Mylar)
	reflectors were used at the open end of the fibre. Results
	for two different discriminator thresholds are shown.}
    \begin{tabular}{lcccc}\\
      \hline
      $V_{thr}$ & open & 1st paint & 2nd paint & Mylar \\
      \hline
      \hline
      30\,mV & 1. & 1.19 & 1.14 & 1.13 \\
      40\,mV & 1. & 1.22 & 1.18 & 1.16 \\
      \hline
    \end{tabular}
  \label{tab:reflectors}
  \end{center}
\end{table}
\begin{table}[tbp]
  \begin{center}
    \caption{ Parameters of double exponential fits to counting rate
      measurements with single and double clad WLS fibres. The second
      set of measurements included an extra-mural absorber ({\it
      EMA}).}
    \begin{tabular}{lcccc}\\
      \hline
      Cladding  & $\ I_L$ & $\lambda_L$\,(cm) & $\ I_S$ 
		& $\lambda_S$\,(cm) \\
      \hline
      \hline
      single    &  9.1$\pm$0.9 & 203$\pm$22 & 14.3$\pm$0.8 
		& 32$\pm$2\\
      double    & 10.8$\pm$0.8 & 289$\pm$24 & 11.3$\pm$0.6 
		& 35$\pm$4\\
      single+{\it EMA} & 7.4$\pm$0.2 & 365$\pm$103 & $1.51\times I_L$
		& 19.1$\pm$2.8\\
      double+{\it EMA} & 9.4$\pm$0.1 & 252$\pm$16 & $1.13\times I_L$
		& 21.2$\pm$1.6\\
      \hline
    \end{tabular}
  \label{tab:attenuation}
  \end{center}
\end{table}
\begin{table}[tbp]
  \begin{center}
    \caption{Measured acceptances, $\alpha$, and global efficiencies,
    $\epsilon$, for cosmic ray muons of the two planes of the
    scintillator hodoscope.}
    \begin{tabular}{lcc}\\
      \hline
      & top & bottom \\

      \hline
      \hline
      $\alpha\, (\%)$   & 96.2 & 96.9 \\
      $\epsilon\, (\%)$ & 98.6 & 98.7 \\
      \hline
    \end{tabular}
  \label{tab:efficiencies}
  \end{center}
\end{table}
%


%
\begin{figure}[tbp]
  \begin{center}
     \epsfig{width= \textwidth, file= 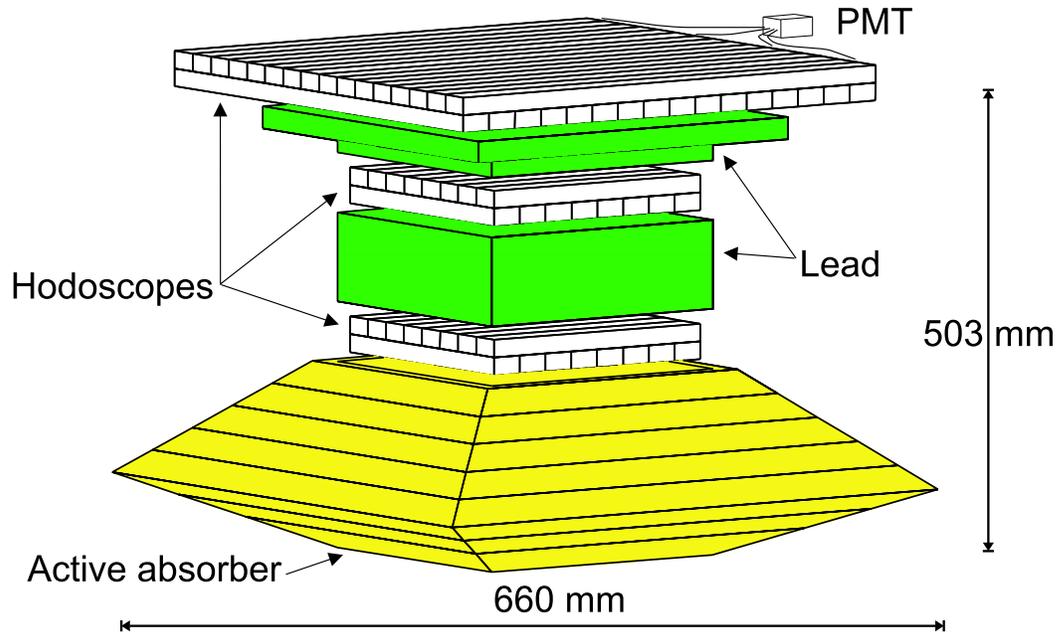}
     \caption{ Sketch of the inner components of the proposed {\sc
       Adler} detector which are relevant for the design of the
       hodoscope. Approximately 100 individual read-out channels from
       three hodoscopes will be needed for muon identification and
       track reconstruction. Six multi-anode PMTs with simple
       discriminator boards attached to their bases are sufficient to
       perform the experiment.}
     \label{Fig:adler_sketch}
  \end{center}
\end{figure}
\begin{figure}[tbp]
  \begin{center}
    \subfigure[]{\epsfig{width= 0.48 \textwidth, 
	file= 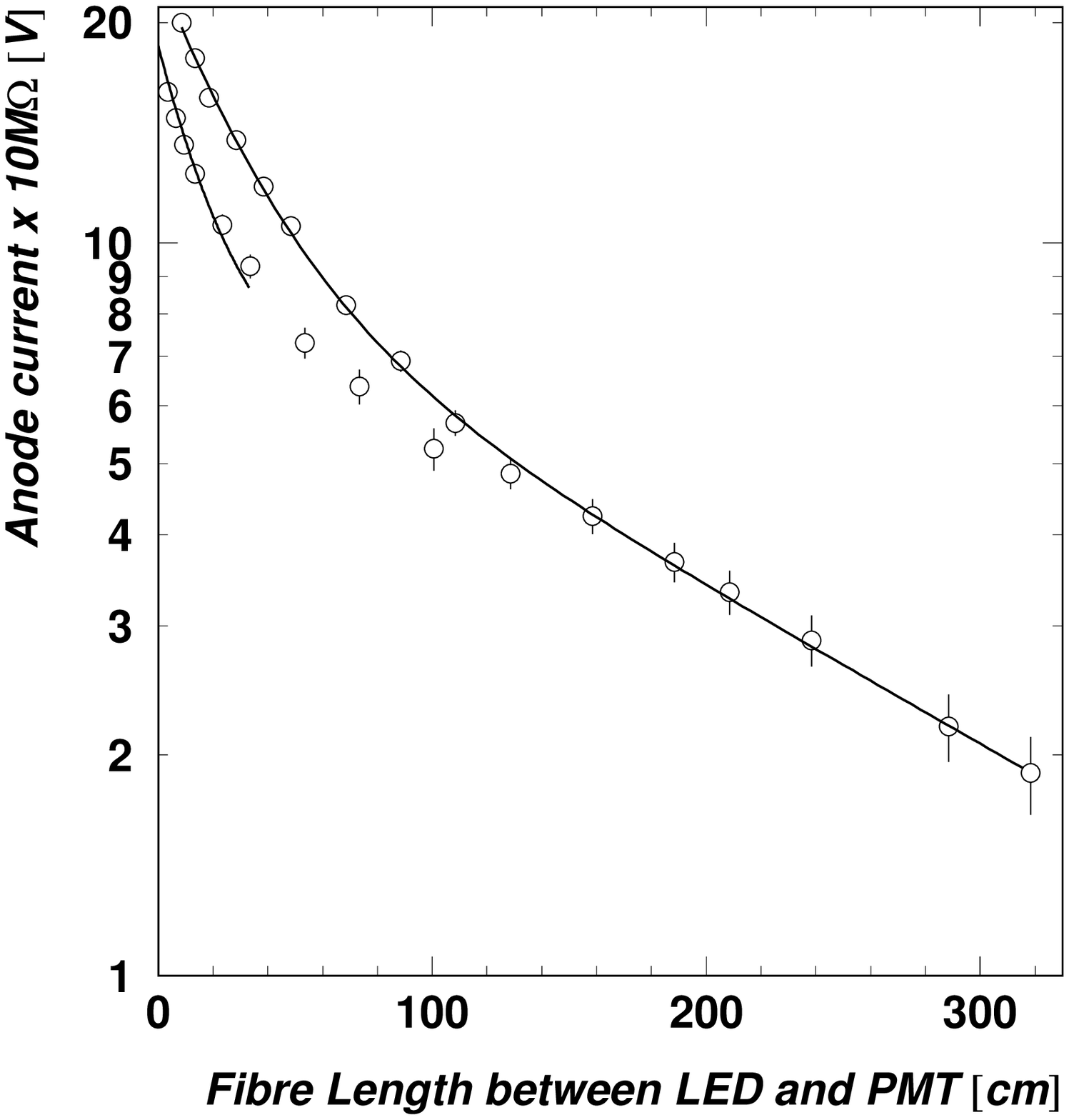}}
    \subfigure[]{\epsfig{width= 0.48 \textwidth, 
	file= 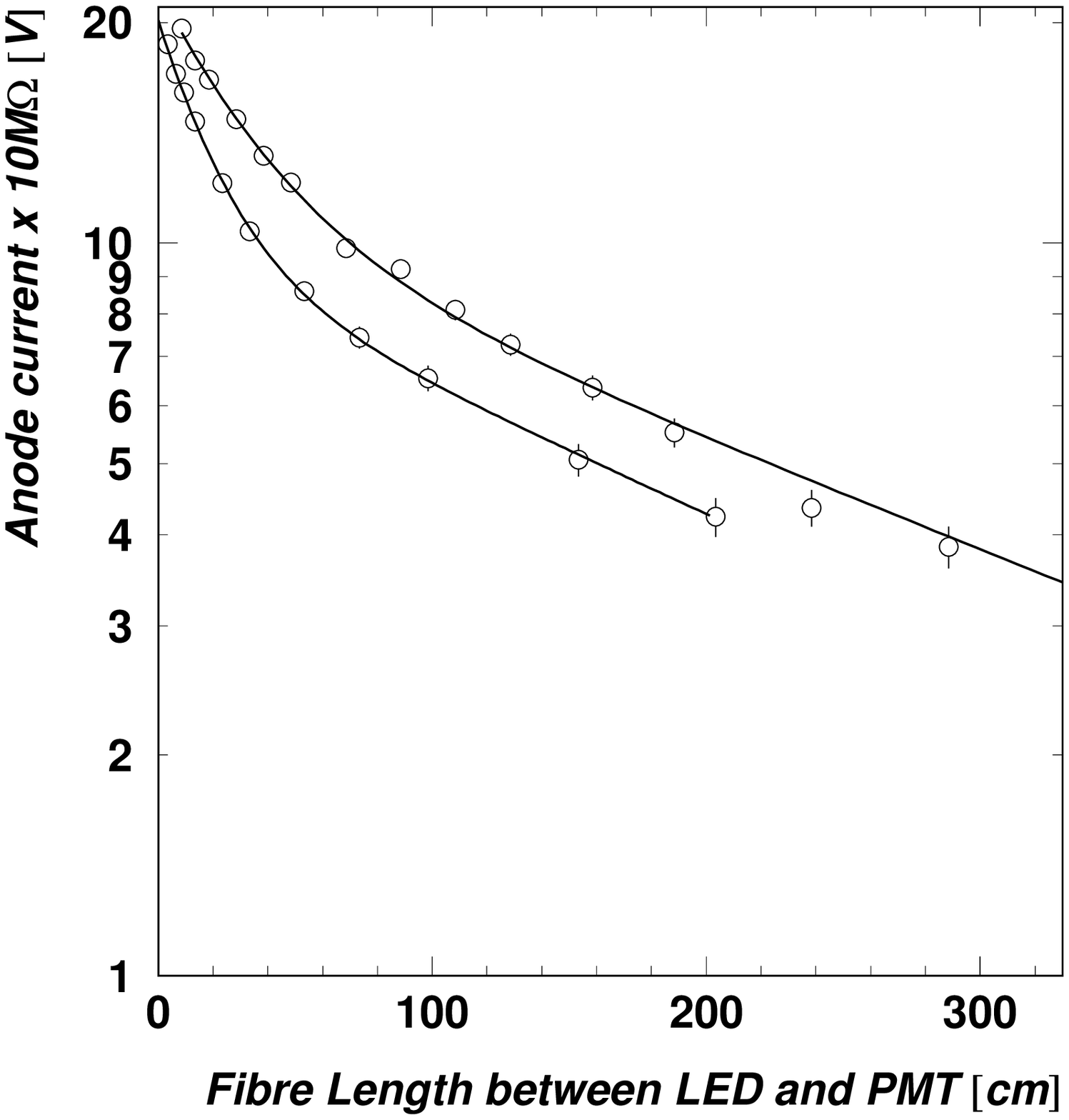}}
    \caption{ Counting rate as a function of WLS fibre length.  The
	plot shows the data points for single (a) and double clad (b)
	BCF-91A fibres. Short and long attenuation lengths were
	determined by double exponential fits. The two lower data sets
	were measured with fibres whose surface were covered by
	extra-mural absorbers ({\it EMAs}).}
    \label{Fig:attenuation}
  \end{center}
\end{figure}
\begin{figure}[tbp]
  \begin{center}
      \epsfig{width= 0.6 \textwidth, file= 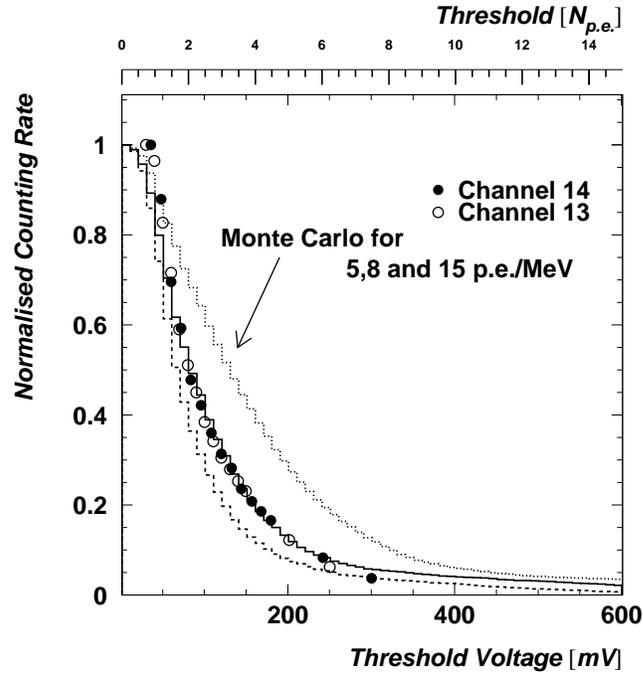}
      \caption{ Normalised counting rate as a function of
	discriminator threshold. The points show two measured PMT
	channels and the histograms show simulations for different
	effective light yields. The best match between measurement and
	simulation is achieved for a light yield of
	8\,photo-electrons$/$MeV.}
      \label{Fig:light-yield}
  \end{center}
\end{figure}
\begin{figure}[tbp]
  \begin{center}
     \epsfig{width= 0.8 \textwidth, file= 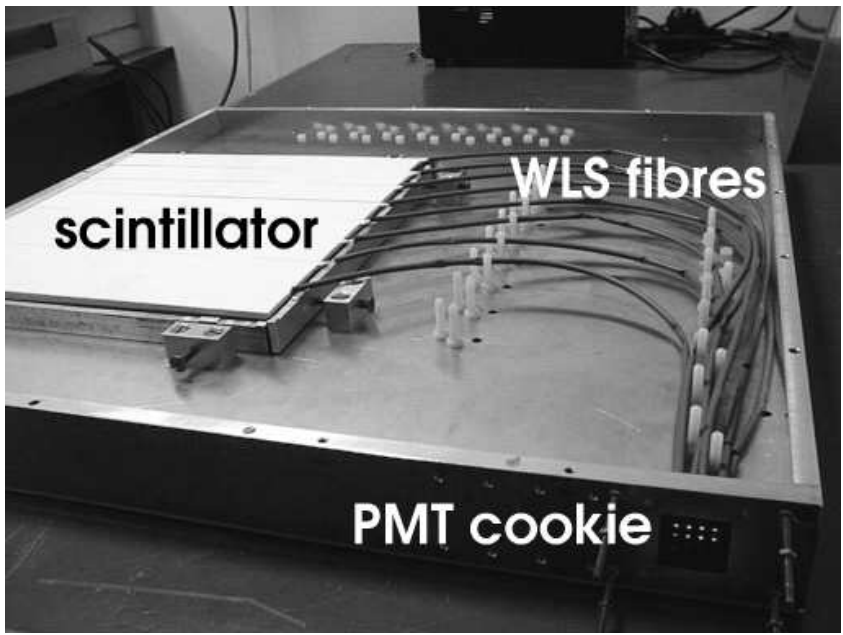}
     \caption{ Internal view of the hodoscope showing the top plane
	scintillators and the routing of the WLS fibres to a cookie
	where a PMT and two discriminator boards can be attached. The
	aluminium frame provided considerable rigidity for relatively
	little weight.}
     \label{Fig:photograph}
  \end{center}
\end{figure}
\begin{figure}[tbp]
  \begin{center}
     \epsfig{width= \textwidth, file= 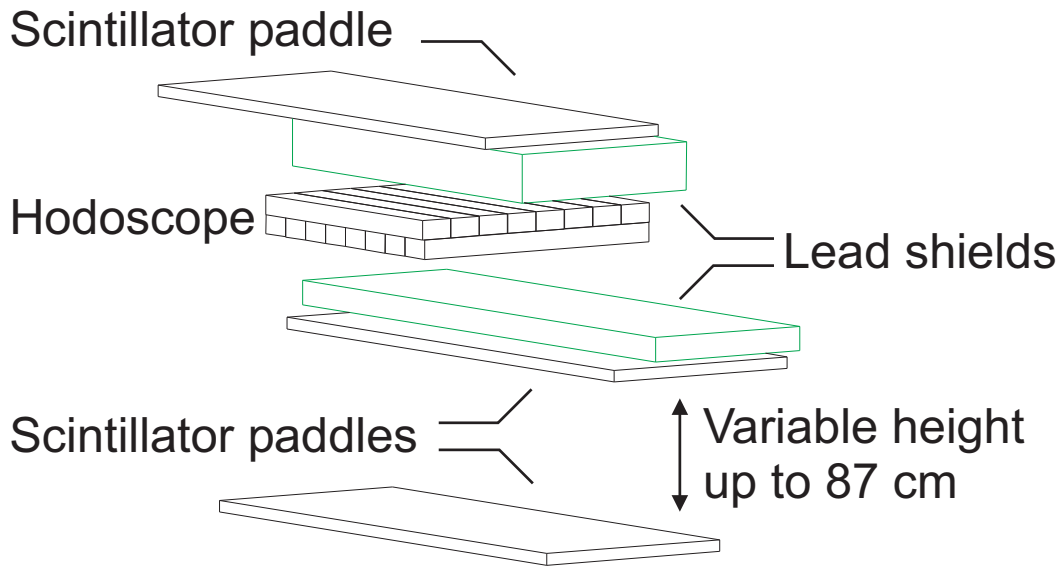}
     \caption{ Sketch of the cosmic ray test stand for the
	hodoscope. Three scintillator paddles and two lead shields
	were used to derive a clean muon trigger.}
     \label{Fig:set-up}
  \end{center}
\end{figure}
\begin{figure}[tbp]
  \begin{center}
     \epsfig{width= 0.8 \textwidth, file= 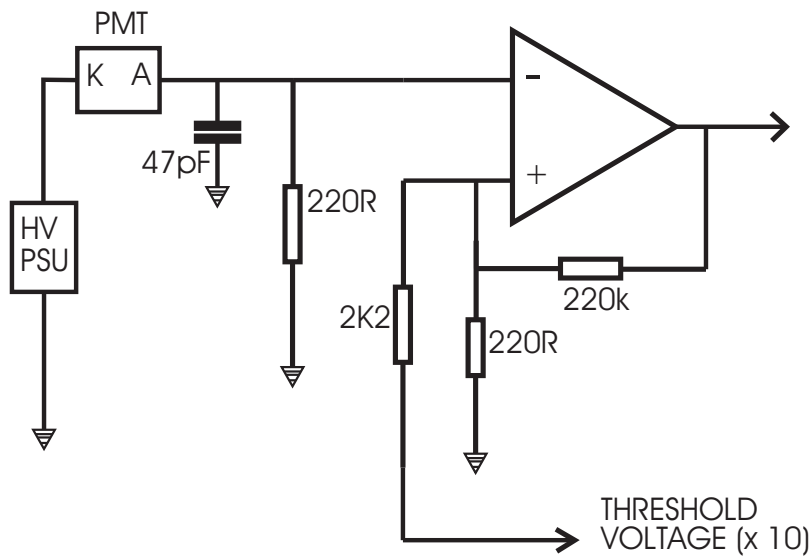}
     \caption{ One channel of the discriminator circuit. Two boards
       carrying eight channels were mounted directly on the base of
       the PMT. A common threshold voltage was provided for each group
       of eight channels.}
     \label{Fig:discr}
  \end{center}
\end{figure}
\begin{figure}[tbp]
  \begin{center}
    \epsfig{width= \textwidth, file=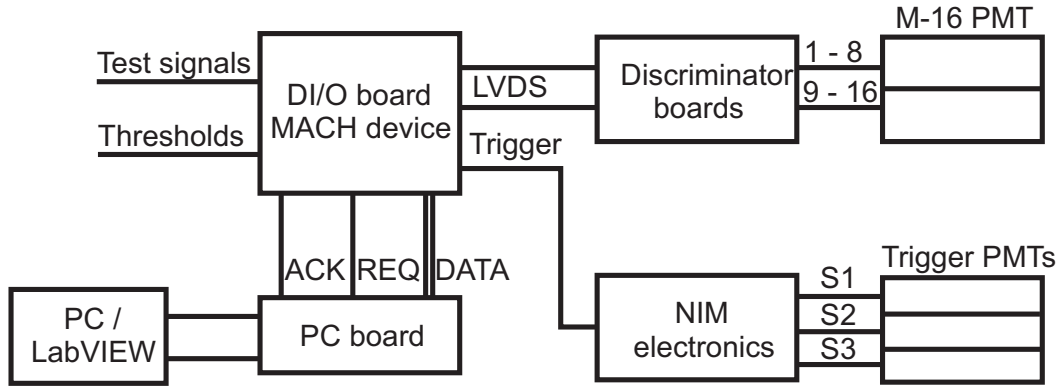}
    \caption{ The data acquisition scheme for the read-out of the
      hodoscope. The signals were latched from the M16 to a PC by a
      state machined programmed `MACH 435' chip. Two 34-way
      flat-n-twist cables carry discriminator output (LVDS), the
      analogue and digital power, test signals and the common
      thresholds. The REQ and ACK lines control the handshaking. The
      signals of the trigger PMTs are discriminated in a NIM unit.}
    \label{Fig:dio}
  \end{center}
\end{figure}
\begin{figure}[tbp]
  \begin{center} 
    \subfigure[]{\epsfig{width= 0.48 \textwidth, file= 
		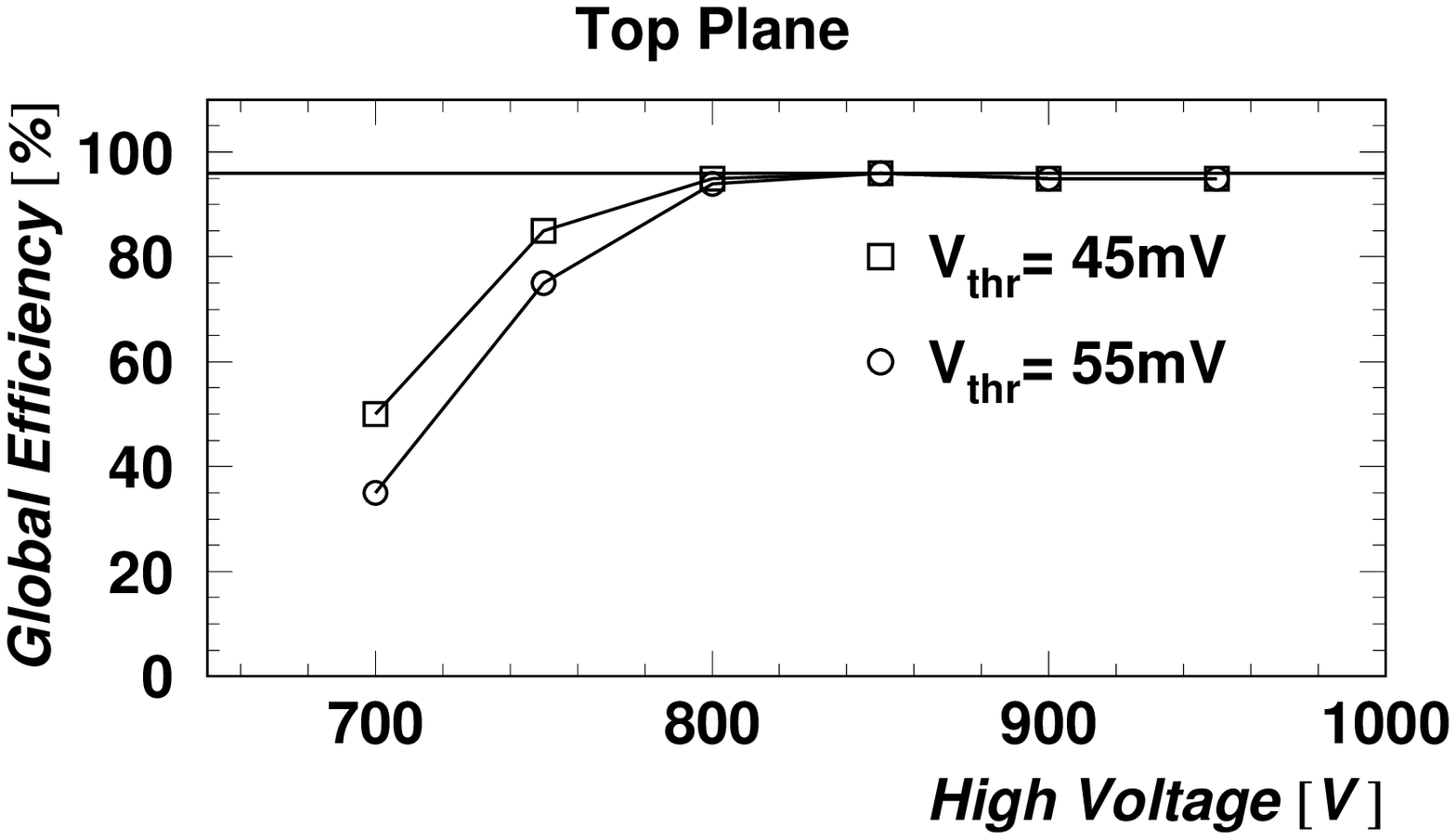}}
    \subfigure[]{\epsfig{width= 0.48 \textwidth, file= 
		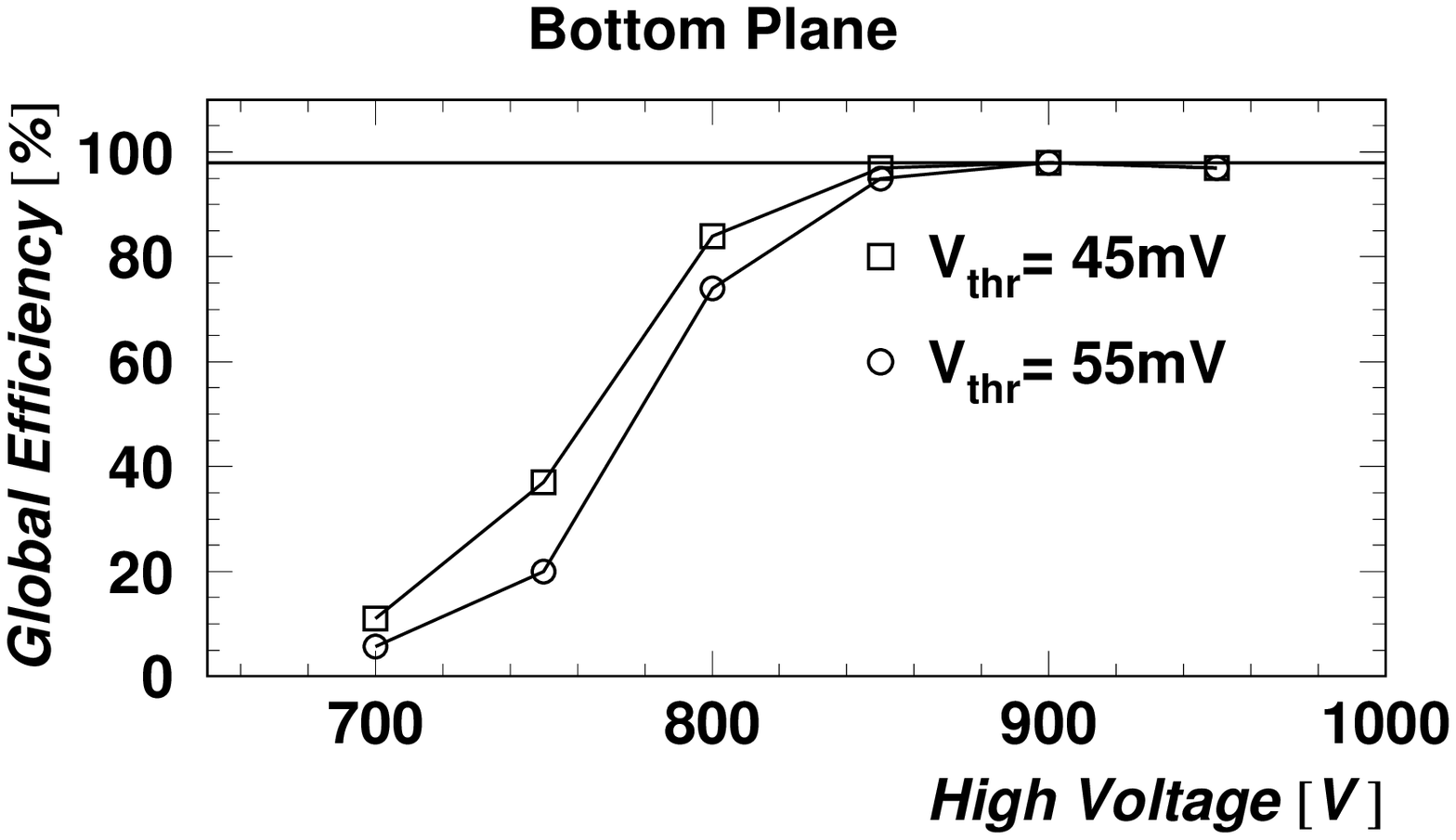}}
    \caption{ Global efficiency of the top (a) and bottom (b)
	hodoscope plane for different HVs and discriminator
	thresholds. The measurements were performed with two
	scintillator paddles in the trigger.}
    \label{Fig:hv_scan} 
  \end{center}
\end{figure}
\begin{figure}[tbp]
  \begin{center} 
    \subfigure[]{\epsfig{width= 0.48 \textwidth, file= 
		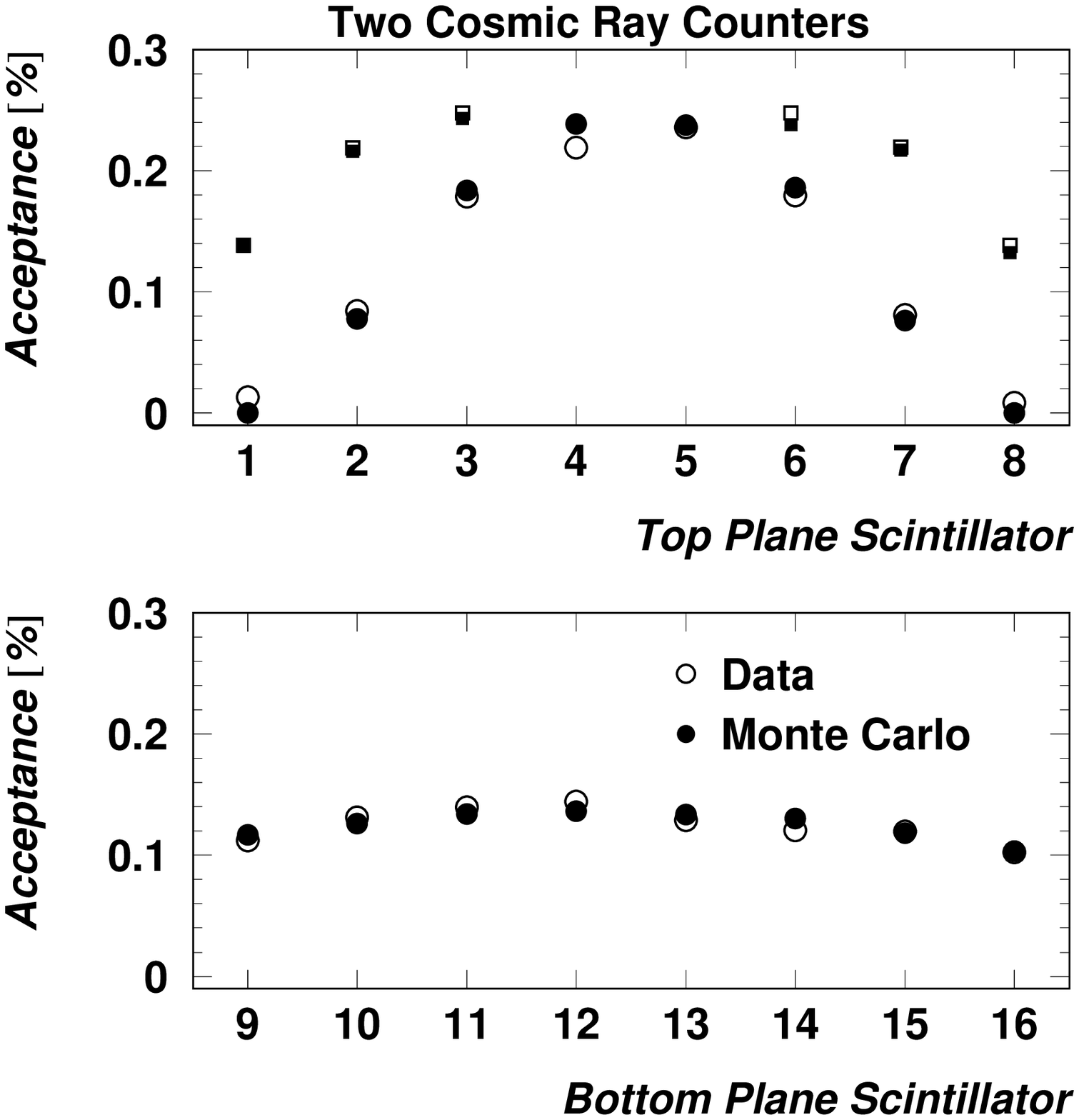}}
    \subfigure[]{\epsfig{width= 0.48 \textwidth, file= 
		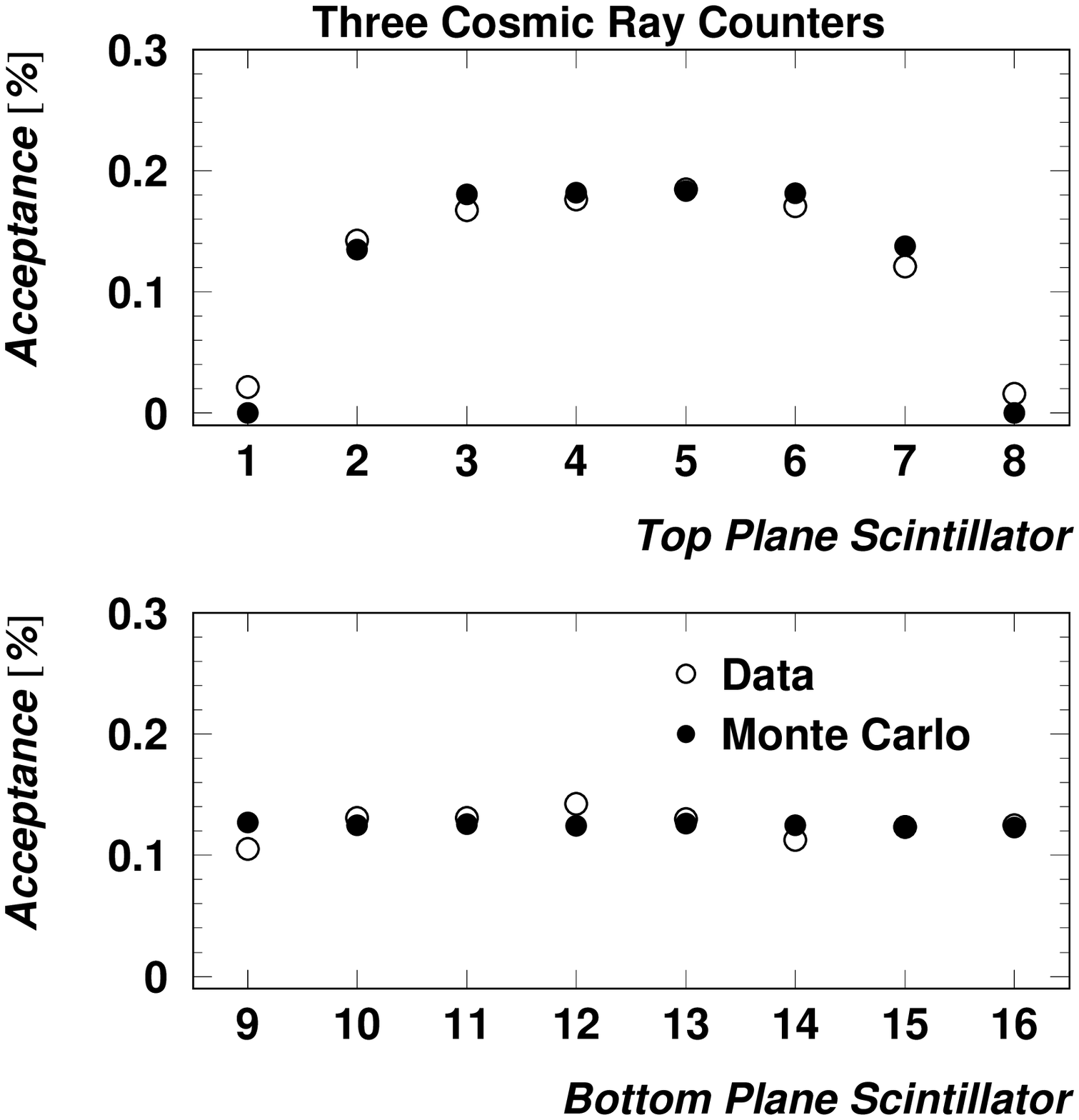}}
    \caption{ Relative acceptance of the individual channels of the
	hodoscope for two scintillator paddles (a) and three
	scintillator paddles (b) in the trigger. Open circles
	represent measurements, full circles represent
	simulation. Squares show the acceptances for the hodoscope
	moved by 5\,cm in either direction.}
    \label{Fig:acceptance} 
  \end{center}
\end{figure}
\begin{figure}[tbp]
  \begin{center} 
    \subfigure[]{\epsfig{width= 0.48 \textwidth, file= 
		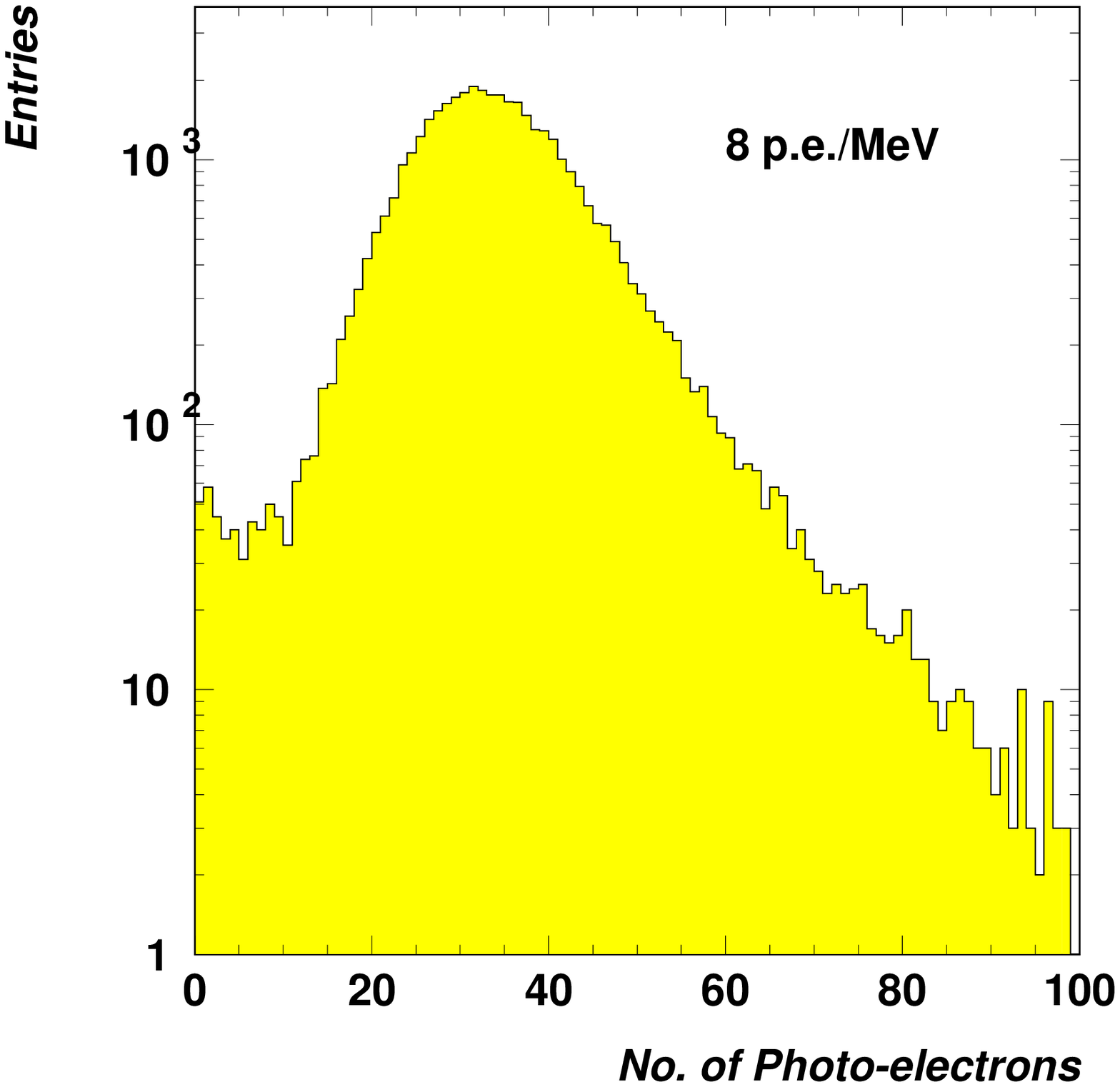}}
    \subfigure[]{\epsfig{width= 0.48 \textwidth, file= 
		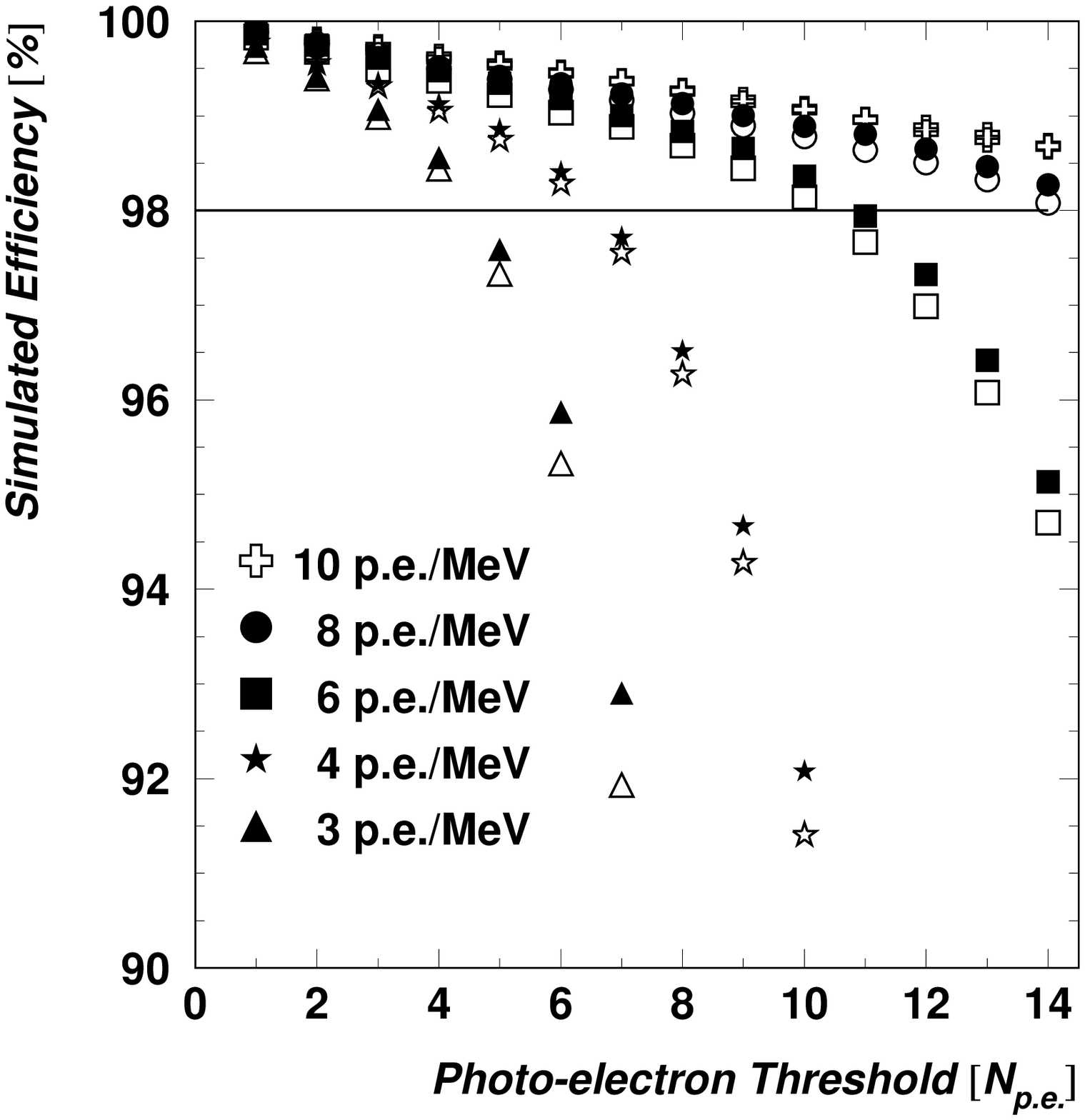}}
    \caption{ Simulated photo-electron spectrum (a) for the
	scintillator bars and the corresponding intrinsic efficiency
	for detecting cosmic ray muons (b). Full and open symbols
	represent top and bottom plane efficiencies, respectively. The
	light yield of the hodoscope was determined from the measured
	efficiency of about 98\,\% (horizontal line).}
    \label{Fig:simulation} 
  \end{center}
\end{figure}
%

\end{document}